\documentclass{article}
\usepackage{spconf,amsmath,graphicx}
\usepackage{amssymb,amsfonts}
\usepackage{textcomp}
\usepackage{xcolor}
\usepackage{booktabs}
\usepackage{graphicx,amsfonts,amscd,amssymb,bm,url,color,latexsym,bbm,amsthm,amsmath}
\usepackage{physics}

\usepackage{algorithm}
\usepackage{algpseudocode}
\usepackage{amsmath, amssymb}
\usepackage{tikz}
\usepackage{xcolor}
\usepackage{graphicx}
\usepackage{subcaption}

\allowdisplaybreaks
\usepackage{hyperref}
\usepackage[capitalize,nameinlink]{cleveref}

\newcommand{\mb}{\boldsymbol}

\newcommand{\mc}{\mathcal}

\newcommand{\bb}{\mathbb}
\newcommand{\set}[1]{\left\{ #1 \right\}}

\newcommand{\eps}{\varepsilon}

\newcommand{\paren}{\pqty}

\DeclareMathOperator*{\argmin}{arg\,min}

\newcommand{\wh}{\widehat}

\numberwithin{equation}{section}

\usepackage{subcaption}
\usepackage{multirow}
\usepackage{adjustbox} 
\usepackage{array}     
\usepackage{rotating}  
\usepackage{balance}

\title{FMPlug: Plug-In Foundation Flow-Matching Priors for Inverse Problems
}

\name{Yuxiang Wan, Ryan Devera, Wenjie Zhang, Ju Sun\thanks{This work was partially supported by NSF ACED 2435911.}}
\address{Department of Computer Science and Engineering, University of Minnesota, Minneapolis, USA\\
Emails: \{wan01530, dever120, zhan7867, jusun\}@umn.edu}
\begin{document}
%
\maketitle
\begin{abstract}
We present FMPlug, a novel plug-in framework that enhances foundation flow-matching (FM) priors for solving ill-posed inverse problems. Unlike traditional approaches that rely on domain-specific or untrained priors, FMPlug smartly leverages two simple but powerful insights: the similarity between observed and desired objects and the Gaussianity of generative flows. By introducing a time-adaptive warm-up strategy and sharp Gaussianity regularization, FMPlug unlocks the true potential of domain-agnostic foundation models. Our method beats state-of-the-art methods that use foundation FM priors by significant margins, on image super-resolution and Gaussian deblurring. 


\end{abstract}
\section{Introduction} \label{sec:intro}
Inverse problems (IPs) are prevalent in many fields, such as medical imaging, remote sensing, and computer vision. IPs aim to recover an unknown object $\mb x$ from its measurement $\mb y \approx \mc A(\mb x)$, where the forward operator $\mc A(\cdot)$ models the measurement process and the approximation sign $\approx$ accounts for potential modeling errors and measurement noise. Due to insufficient measurement and/or the approximate relationship in $\mb y \approx \mc A(\mb x)$, in practice $\mb x$ is typically not uniquely recoverable from $\mb y$ alone, i.e., ill-posedness. So, to obtain a reliable and meaningful solution for IPs, it is important to incorporate prior knowledge of $\mb x$. 

Traditional ideas for solving IPs rely on optimization formulations, often motivated under the Maximum A Posteriori (MAP) estimation principle: 
\begin{align}     \label{eq:inv_sol}
    \min\nolimits_{\mb x} \; \ell (\mb y, \mc A (\mb x)) + \Omega (\mb x).     
\end{align}
Here, minimizing data fitting loss $\ell(\mb y, \mc A(\mb x))$ encourages $\mb y \approx \mc A(\mb x)$, and the regularization term $\Omega(\mb x)$ encodes prior knowledge of ideal solutions to resolve ambiguities and hence mitigate potential ill-posedness. The optimization problems are often solved by gradient-based iterative methods. Advances in deep learning (DL) have revolutionized IP solving. Different DL-based approaches to IPs operate with different data-knowledge tradeoffs. For example, supervised approaches take paired datasets $\set{\mb y_i, \mb x_i}_{i=1}^N$ and directly learn the inverse mapping $\mb y \mapsto \mb x$, with or without using $\mc A$~\cite{Ongie2020DeepLearning,monga2021algorithm,zhang2024wrong}; alternatively, data-driven priors learned from object-only datasets $\set{\mb x_i}_{i=1}^N$ can be integrated with \cref{eq:inv_sol}~\cite{oliviero2025generative, daras2024surveydiffusionmodelsinverse,Wang2024DMPlugAP}; strikingly, untrained DL models themselves can serve as plug-in priors for \cref{eq:inv_sol}, without any extra data~\cite{alkhouri2025understanding,wang2023early,li_deep_2023,zhuang_blind_2023,zhuang_practical_2023,li2021self}. \cite{Ongie2020DeepLearning} give comprehensive reviews of these ideas. 

In this paper, we focus on solving IPs with pretrained deep generative priors (DGPs)~\cite{oliviero2025generative}. Compared to supervised approaches that perform task-specific training, these approaches enjoy great flexibility, as DGPs can be plugged in and reused for different tasks related to the same family of objects. Among the different DGPs, \textbf{we are most interested in those based on diffusion models (DMs), and the more general flow-matching (FM) models~\cite{lipman2024flow}}, as they are the backbone of state-of-the-art (SOTA) generative models in several domains~\cite{labs2025flux1kontextflowmatching,esser2024scalingrectifiedflowtransformers,Agarwal2025Cosmos}.  

Numerous recent works target solving IPs with pretrained DM/FM priors~\cite{daras2024surveydiffusionmodelsinverse}. Although promising, most of them are based on \textbf{domain-specific} DM/FM priors, e.g., trained on the \texttt{FFHQ} dataset for human faces and the \texttt{LSUN bedrooms} dataset for bedroom scenes. This limits the practicality of these methods, as domain-specific DM/FM models are not always readily available. On the other hand, the emergence of domain-agnostic \textbf{foundation} DM/FM models, such as Stable Diffusion~\cite{esser2024scalingrectifiedflowtransformers} and Flux~\cite{labs2025flux1kontextflowmatching} for images, seems to obsolete domain-specific developments; \cite{kim2025flowdpsflowdrivenposteriorsampling, patel2024steering,ben2024d} propose such ideas. \textbf{However, the reported performance clearly lags behind those obtained with domain-specific FM/DM priors, and even behind those obtained with untrained priors}; see \cref{sec:ffm-ips-related}. This should not be surprising, as in terms of solving any particular IP, foundation priors are considerably weaker than domain-specific ones and not clearly stronger than untrained priors. 

In this paper, we describe our ongoing efforts to close the performance gap. We focus on image restoration and take advantage of two things to strengthen the image prior: (1) the similarity of $\mb x$ and $\mb y$, which motivates us to develop a learnable warm-up strategy; and (2) Gaussianity in DM/FM priors, suggesting Gaussianity regularization. By incorporating these two simple changes, we demonstrate convincing performance gains over prior work based on foundation FM priors.

\section{Technical background \& related Work}
\subsection{Flow matching (FM)}

Flow Matching (FM) models are an emerging class of deep generative models~\cite{lipman2024flow}. They learn a continuous flow to transform a prior distribution $p_0(\mb z)$ into a target distribution $p_1(\mb x)$---in the same spirit of continuous normalizing flow (CNF), described by an ordinary differential equation (ODE)
\begin{align}
    d\mb z  = \mb v(\mb z, t)\ dt. 
\end{align}
Whereas CNF focuses on the density path induced by the flow and performs the maximum likelihood estimation, FM tries to learn a parametrized velocity field $\mb v_{\mb \theta}(\mb z, t)$ to match the one associated with the desired flow. To generate new samples after training, one simply samples $\mb z_0 \sim p_0(\mb z)$ and numerically solves the learned ODE induced by $\mb v_{\mb \theta}(\mb z, t)$ from $t=0$ to $t=1$, to produce a sample $\mb z_1 \sim p_1(\mb x)$. 

For tractability, in practice, FM matches the conditional velocity field: for each training point $\mb x$, a simple conditional probability path $p_t(\mb{z}| \mb{x})$, e.g., induced by a linear interpolation $\mb z_t =  t \mb x + (1-t) \mb z_0$, is defined. The model is then trained to learn the known vector field of this conditional path
\begin{align}
    \min\nolimits_{\mb \theta} \; \mathbb{E}_{\mb x, \mb z_0, t} \left\| \mb v_\theta(\mb z_t, t) - \mb {u}(\mb {z}_t, t | \mb {x}) \right\|^2. 
\end{align}

DMs based on probability flow ODEs can also be interpreted as FMs, although they match the score functions induced by the chosen probability path, not the vector field~\cite{lipman2024flow}. So, FM can be viewed as a general generative framework that covers DM as well besides other algorithmic possibilities. 

\subsection{Pretrained FM priors for IPs}
Recent methods that use pretrained FM priors for solving IPs can be classified into two families: \textbf{(1) The interleaving approach} interleaves the generation steps (i.e., numerical integration steps) with gradient steps toward feasibility (i.e., making $\mb x$ satisfies that $\mb y \approx \mc A(\mb x)$)~\cite{pokle2023training, kim2025flowdpsflowdrivenposteriorsampling, patel2024steering,martin2025pnpflowplugandplayimagerestoration}; see a sample algorithm in \cref{alg:unified_interleaving}. Despite the simplicity and empirical effectiveness on simple IPs, these methods might not converge or return an $\mb x$ that respects the data prior (i.e., \textbf{manifold feasibility}) and/or satisfies the measurement constraint $\mb y \approx \mc A(\mb x)$ (i.e., \textbf{measurement feasibility}); and \textbf{(2) the plug-in approach} views the entire generation process (lines 4--8 in \cref{alg:plugin_optimization}) as a function $\mc G_{\mb \theta}$ that maps any source sample to a target sample, and plugs the prior into \cref{eq:inv_sol} to obtain a unified optimization formulation~\cite{ben2024d, Wang2024DMPlugAP,wang2025temporal} ($\circ$: function composition): 
{\small 
\begin{align} \label{eq:ours_fm}
\mb z^\ast \in \argmin\nolimits_{\mb z} \; \mc L(\mb z) \doteq  \ell\paren{\mb y, \mc A \circ \mc G_{\mb \theta}(\mb z)} + \Omega \circ \mc G_{\mb \theta}(\mb z);
\end{align} 
}
see \cref{alg:plugin_optimization} for a sample algorithm. The estimated object is then $\mc G_{\mb \theta}(\mb z^\ast)$. Here, the generator $\mc G_{\mb \theta}$ is fixed and the output $\mc G_{\mb \theta}(\mb z)$ naturally satisfies the manifold feasibility. In addition, global optimization of $\mc L(\mb z)$ forces small $\ell\paren{\mb y, \mc A \circ \mc G_{\mb \theta}(\mb z)}$, and hence $\mb y \approx \mc A \circ \mc G_{\mb \theta}(\mb z)$, i.e., measurement feasibility. 
\begin{algorithm}[H]
\centering 
\caption{A sample algorithm of the interleaving approach}
\label{alg:unified_interleaving}
\begin{algorithmic}[1]
\Require ODE steps $T$, measurement $\mb y$, forward model $\mc A$
\State Initialize $\mb z_0 \sim \mc N(\mb 0, \mb I)$
\For{$i = 0$ to $T-1$}
    \State $t_i \gets i / T$
    \State $\mb {v} \gets \mb{v}_{\mb \theta}(\mb z_i, t_i)$ \Comment{learned velocity}
    \State $\mb z'_{i+1} \gets \mb z_i + 1/T \cdot \mb{v} $ \Comment{discrete integration}
    \State $\mb z_{i+1} \gets$ $(\mb y, \mc A)$-driven update of $\mb z'_{i+1}$ \Comment{reducing $\ell(\mb y, \mc A(\mb z))$ starting from $\mb z'_{i+1}$}
\EndFor
\Ensure Estimated $\wh{\mb x}$
\end{algorithmic}
\end{algorithm}
\vspace{-1em}
\begin{algorithm}[H]
\centering
\caption{A sample algorithm of the plug-in approach}
\label{alg:plugin_optimization}
\begin{algorithmic}[1]
\Require Total iterations $E$, ODE steps $T$, measurement $\mb y$, forward model $\mc A$ 
\State Initialize $\mb z^{(0)} \sim \mc N(\mb 0, \mb I)$
\For{$e = 0$ to $E - 1$}
    \State $\mb z_0 \gets \mb z^{(e)}$ 
    \For{$i = 0$ to $T-1$}  \Comment{whole integration path}
        \State $t_i \gets i / T$
        \State $\mb{v} \gets \mb{v}_{\mb \theta}(\mb z_i, t_i)$
        \State $\mb z_{i+1} \gets \mb z_i + 1/T \cdot \mb{v}$
    \EndFor
    \State $\mb z^{(e+1)} \gets \mb z^{(e)} - \eta^{(e)} \cdot \nabla_{\mb z} \mc L(\mb z_{i+1})$  \Comment{reducing loss}
\EndFor
\Ensure Estimation $\wh{\mb x} = \mc G_{\mb \theta}(\mb z^{(E-1)})$
\end{algorithmic}
\end{algorithm}


\subsection{Foundation FM priors for IPs}
\label{sec:ffm-ips-related}
\begin{table}[!htpb]
\vspace{-1em}
\caption{A comparison between foundation, domain-specific, and untrained priors for Gaussian deblurring on AFHQ-Cat (resolution: $256 \times 256$; DS: domain-specific; FD: foundation). \textbf{Bold}: best, \& \underbar{underline}: second best, for each metric/column. }
\vspace{-1.5em}
\label{tab:generic_vs_domain}
\begin{center}
\resizebox{0.8\linewidth}{!}{  
\setlength{\tabcolsep}{1.0mm}{
\begin{tabular}{c c c c c}
\hline
&\scriptsize{PSNR$\uparrow$}
&\scriptsize{SSIM$\uparrow$}
&\scriptsize{LPIPS$\downarrow$}
&\scriptsize{CLIPIQA$\uparrow$}
\\
\hline
\scriptsize{\textbf{DIP}}
&\scriptsize{\underbar{27.5854}}
&\scriptsize{\underbar{0.7179}}
&\scriptsize{0.3898}
&\scriptsize{0.2396}
\\
\hline
\scriptsize{\textbf{D-Flow (DS)}}
&\scriptsize{\textbf{28.1389}}
&\scriptsize{\textbf{0.7628}}
&\scriptsize{\textbf{0.2783}}
&\scriptsize{\textbf{0.5871}}
\\
\hline
\scriptsize{\textbf{D-Flow (FD)}}
&\scriptsize{25.1453 }
&\scriptsize{0.6829 }
&\scriptsize{0.5213}
&\scriptsize{0.3228}
\\
\hline 
\scriptsize{\textbf{FlowDPS (DS)}}
&\scriptsize{22.1191}
&\scriptsize{0.5603}
&\scriptsize{\underbar{0.3850}}
&\scriptsize{\underbar{0.5417}}
\\
\hline
\scriptsize{\textbf{FlowDPS (FD)}}
&\scriptsize{22.1404 }
&\scriptsize{0.5930}
&\scriptsize{0.5412 }
&\scriptsize{0.2906 }
\\
\hline
\end{tabular}
}
}
\end{center}
\vspace{-1.5em}
\end{table}
\noindent The availability of large-scale training sets has recently fueled intensive development of foundation generative models in several domains, most of them based on FM models and variants, e.g., Stable Diffusion V3 (and newer)~\cite{esser2024scalingrectifiedflowtransformers} and FLUX.1~\cite{labs2025flux1kontextflowmatching} for images, OpenAI Sora~\cite{OpenAISora2024} and Google Veo~\cite{Veo_DeepMind_2025} for videos, and Nvidia Cosmos world model~\cite{Agarwal2025Cosmos}. By contrast, domain-specific FM models are not always readily available (e.g., due to the lack of training data for scientific applications). So, recent IP methods based on pretrained FM priors have started to explore foundation priors. 

However, while these foundation priors only constrain the object to be physically meaningful (e.g., the object being a natural image), 
domain-specific priors provide much more semantic and perhaps structural information about the object (e.g., the object being a facial or brain MRI image), i.e., \textbf{foundation priors alone are considerably weaker than domain-specific priors}. In fact, untrained priors such as deep image priors may be powerful enough to promote physically meaningful reconstruction~\cite{alkhouri2025understanding,wang2023early,li_deep_2023,zhuang_blind_2023,zhuang_practical_2023}. A quick comparison summarized in \cref{tab:generic_vs_domain} confirms our intuition: \textbf{recent IP methods with foundation priors perform worst compared to domain-specific and even untrained priors}.   

To strengthen the object prior, \cite{ben2024d} assumes similarity of $\mb x$ and $\mb y$---valid for typical image restoration problems, and initializes \cref{eq:ours_fm} with  
\begin{align}  \label{eq:d-flow-init}
    \mb z_0 = \sqrt{\alpha} \mb y_0 + \sqrt{1-\alpha} \mb z  \quad \text{where} \; \mb z \sim \mc N(\mb 0, \mb I), 
\end{align}
where $\mb y_0$ is the backward solution of the governing ODE, i.e., $\mb y_0 = \mb y + \int_{1}^0 \mb v_{\mb \theta}(\mb y_t, t) dt$, or inverted seed in other words. Moreover, they promote the Gaussianity of the seed $\mb z_0$ by recognizing that $\norm{\mb z_0}_2^2$ follows a $\chi^2$ distribution and thus regularizes its negative log-likelihood; \cite{kim2025flowdpsflowdrivenposteriorsampling} takes automatically generated text prompts for $\mb y$ from another model as text conditions, as all recent foundation FM models allow text-prompted generation. 

\section{Method} \label{sec:method}
We focus on image restoration with foundation FM priors, and aim to strengthen the image prior based on (1) the similarity of $\mb x$ and $\mb y$, and (2) Gaussianity in FM priors, in the same vein as \cite{ben2024d} but improving upon theirs.

\subsection{Time-dependent warm-up strategy} \label{sec:warm-up}

In the standard FM setting, the source distribution $\mb z_0 \sim \mc N(\mb 0, \mb I)$, whereas the initialized $\mb z_0$ in \cref{eq:d-flow-init} has a distribution $\mc N(\sqrt{\alpha} \mb y_0, (1-\alpha) \mb I)$. One might not worry about this distribution mismatch, as both are supported on the entire ambient space anyway. But finite-sample training in practice causes the gap: due to the famous concentration of measure phenomenon~\cite{vershynin2018high}, virtually all training samples drawn from $\mc N(\mb 0, \mb I)$ come from an ultra-thin shell around $\bb S^{d-1}(\mb 0, \sqrt{d})$, a sphere centered at $\mb 0$ and with a radius $\sqrt{d}$: $\bb P [|\norm{\mb z}_2 - \sqrt{d}| \ge t] \le 2e^{-ct^2}$ for a universal constant $c > 0$. If we write that ultra-thin shell spanned by the finite training set as $\mc S$, the generation function $\mc G_{\mb \theta}$ is effectively trained on inputs from the domain $\mc S$, not the entire ambient space: the behavior of $\mc G_{\mb \theta}$ on $\mc S^c$ is largely undetermined. Now, samples from $\mc N(\sqrt{\alpha} \mb y_0, (1-\alpha) \mb I)$ concentrate around another ultra-thin shell around $\bb S^{d-1}(\sqrt{\alpha} \mb y_0, \sqrt{(1-\alpha)d})$, which has only a negligibly small intersection with $\mc S$. So, the initialization in \cref{eq:d-flow-init} lies in $\mc S^c$ with a very high probability. Given that the behavior of $\mc G_{\mb \theta}$ can be wild on $\mc S^c$, this initialization strategy seems problematic. 

We take a different route. We know that a typical flow of FM models takes the form
\begin{align}
    \mb z_t = \alpha_t \mb x + \beta_t \mb z \quad \text{where}\; \mb z \sim \mc N(\mb 0, \mb I), 
\end{align}
where $\alpha_t$ and $\beta_t$ are known functions of $t$ with the property that 
\begin{align}
    \begin{cases}
    \alpha_t {\scriptstyle\searrow} 0, \beta_t {\scriptstyle\nearrow} 1 \; \text{as} \; t \to 0\\    
    \alpha_t {\scriptstyle\nearrow} 1, \beta_t {\scriptstyle\searrow} 0 \; \text{as} \; t \to 1, 
    \end{cases}
\end{align}
where $\scriptstyle \nearrow$ and $\scriptstyle\searrow$ indicate monotonically increasing and decreasing, respectively. Now, when $\mb x$ and $\mb y$ are close, $\mb x = \mb y + \mb \eps$ for some small (i.e., $\norm{\mb \eps}$ is small compared to $\norm{\mb x}$ and $\norm{\mb z}$) but unknown $\mb \eps$. So, we can write the flow as 
\begin{align} \label{eq:exact-flow-y}
    \mb z_t = \alpha_t (\mb y + \mb \eps)  + \beta_t \mb z = \alpha_t \mb y + \beta_t \mb z + \alpha_t \mb \eps \notag \\
    \text{where}\; \mb z \sim \mc N(\mb 0, \mb I). 
\end{align}
To eliminate the unknown $\mb \eps$, we can approximate the exact flow in \cref{eq:exact-flow-y} with the following \textbf{approximate flow} 
\begin{align} \label{eq:approx-flow}
    \mb z_t \approx \alpha_t \mb y + \beta_t \mb z \quad \text{where}\; \mb z \sim \mc N(\mb 0, \mb I)
\end{align}
with an approximation error $\alpha_t \mb \eps$. To control the error, (1) if $\mb \eps$ is relatively large, a small $\alpha_t$ is desirable; (2) if $\mb \eps$ is already relatively small, a relatively large $\alpha_t$ is acceptable. So, although we do not know $\mb \eps$ itself and hence its magnitude, with appropriate $\alpha_t$ we can always make $\alpha_t \mb \eps$ sufficiently small. So, we leave $\alpha_t$ as a learnable parameter. Since $\alpha_t$ is a known function of $t$, we simply need to leave $t \in [0, 1]$ learnable, leading to our warm-start formulation 
\begin{align}\label{eq:warm}
    \min\nolimits_{\mb z, t \in [0, 1]}\;  \ell\paren{\mb y, \mc A \circ \mc G_{\mb \theta} \paren{\alpha_t \mb y + \beta_t \mb z, t}}. 
\end{align}

Note that here we overload the notation of $\mc G_{\mb \theta}$ as $\mc G_{\mb \theta}: \bb R^d \times [0, 1] \to \bb R^d$---the second input is the current $t$ on the path (the notation we use in \cref{eq:ours_fm} assumes $t = 0$). In other words, due to the closeness of $\mb x$ and $\mb y$, we do not need to start from scratch, i.e., a random sample from the source distribution; instead, we plug $\mb y$ into an appropriate---to be learned---time point of the flow to create a shortcut. 

Our formulation in \cref{eq:warm} can be easily generalized to latent FM models. Moreover, it is not only grounded in theory and effective in practice (see \cref{sec:experiment}), but also speeds up learning as $t > 0$ implies shorter flows---although improving the speed is not our current focus. 

Due to the approximation errors in matching the ideal flow during FM training, as well as when approximating \cref{eq:exact-flow-y} using \cref{eq:approx-flow}, the distribution of $\mb z_t$ could be slightly off the ideal distribution. To rectify this, we perform a lightweight variance calibration in our implementation: 
\begin{equation}
    \label{eq:calibration}
    \hat{\mb z}_{t}=\sqrt{\mathrm{Var}(Z_{t})/\mathrm{Var}(\mb z_{t})} \cdot \mb z_{t},
\end{equation} 
where $\mathrm{Var}(Z_{t})$ is the scalar-variable variance across all dimensions and the generation paths of $512$ samples, and $\mathrm{Var}(\mb z_{t})$ scalar-variable variance for $\mb z_t$ across all dimensions. 

\subsection{Gaussianity regularization} \label{sec:reg}
\begin{figure}[!htbp]
    \centering
    \includegraphics[width=0.9\linewidth]{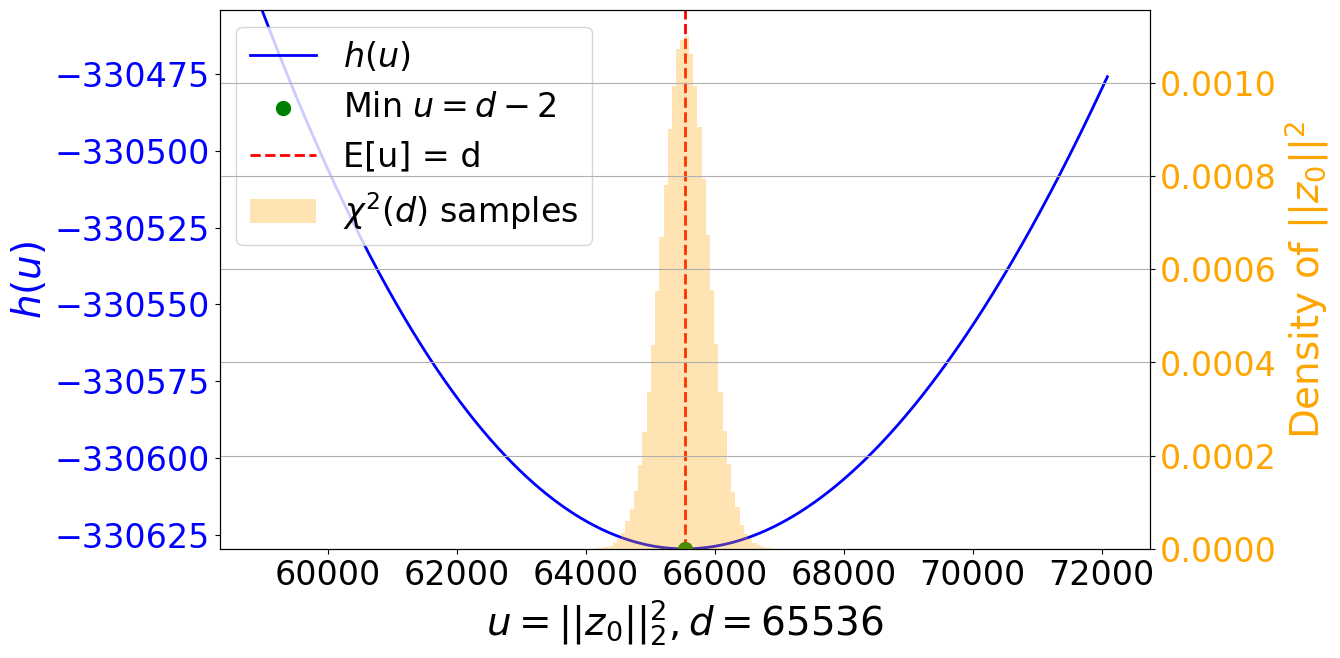}
    \vspace{-1em}
    \caption{Plot of the function $h(\mb z_0)$ (after a change of variable $u = \norm{\mb z_0}_2^2$). An ideal regularization function should blow up sharply away from the narrow concentration region in orange to promote Gaussianity effectively.}
    \label{fig:chi2LL}
    \vspace{-1em}
\end{figure}
If $\mb z_0 \sim \mc N(\mb 0, \mb I)$, $\norm{\mb z_0}_2^2 \sim \chi^2 (d)$ and the negative log-likelihood is $h(\mb z_0) = -(d/2-1) \log \norm{\mb z_0}_2^2 + \norm{\mb z_0}_2^2/2 + C$ for some constant $C$ independent of $\mb z_0$. \cite{ben2024d} promotes the Gaussianity of $\mb z_0$ by regularizing $h(\mb z_0)$. While $h(\mb z_0)$ is minimized at any $\mb z_0$ satisfies $\norm{\mb z_0}_2^2 = d-2$, away from this value the function changes slowly; see \cref{fig:chi2LL}. This is problematic, as $\norm{\mb z_0}_2^2$ should concentrate sharply around $d$ due to concentration of measure and thus only functions that blow up quickly away from the $\norm{\mb z_0}_2^2 = d$ level can effectively promote the Gaussianity of $\mb z_0$.

For our formulation in \cref{eq:warm}, we hope to promote the Gaussianity of $\mb z$. To enforce the sharp concentration of $\mb z$, we project $\mb z$ onto the sphere $\bb S^{d-1}(\mb 0, \sqrt{d})$ in each step by 
\begin{equation} \label{eq:gauss-proj}
    \mb z^\prime = \sqrt{d} \mb z/\| \mb z\|_2. 
\end{equation}
We are not the first to use the more precise spherical constraint or regularization in promoting Gaussianity in the DM/FM literature; see, e.g., \cite{yang2024guidancesphericalgaussianconstraint}.

\section{Experiment}
\label{sec:experiment}



\begin{table}[!htpb]
\vspace{-1em}
\caption{\textbf{x4 Super Resolution} with additive Gaussian noise ($\sigma = 0.03$). (\textbf{Bold}: best, \underbar{underline}: second best)}
\vspace{-1.5em}
\label{tab:sr4}
\begin{center}
\setlength{\tabcolsep}{1.0mm}{
\resizebox{\linewidth}{!}{
\begin{tabular}{c c c c c c c c c}
\hline
&\multicolumn{4}{c}{\scriptsize{\textbf{DIV2K~\cite{Agustsson_2017_CVPR_Workshops}}}}
&\multicolumn{4}{c}{\scriptsize{\textbf{AFHQ~\cite{choi2020starganv2}}}}
\\
\cmidrule(lr){2-5}
\cmidrule(lr){6-9}
&\scriptsize{PSNR$\uparrow$}
&\scriptsize{SSIM$\uparrow$}
&\scriptsize{LPIPS$\downarrow$}
&\scriptsize{CLIPIQA$\uparrow$}

&\scriptsize{PSNR$\uparrow$}
&\scriptsize{SSIM$\uparrow$}
&\scriptsize{LPIPS$\downarrow$}
&\scriptsize{CLIPIQA$\uparrow$}
\\
\hline
\scriptsize{\textbf{DIP}} 

&\scriptsize{24.896}
&\scriptsize{0.613}
&\scriptsize{0.482}
&\scriptsize{0.305}

&\scriptsize{28.275}
&\scriptsize{0.697}
&\scriptsize{0.466}
&\scriptsize{0.284}

\\
\hline
\scriptsize{\textbf{FlowChef}} 

&\scriptsize{24.439}
&\scriptsize{0.677}
&\scriptsize{0.444}
&\scriptsize{\textbf{0.630}}

&\scriptsize{28.444}
&\scriptsize{0.779}
&\scriptsize{0.404}
&\scriptsize{\textbf{0.568}}

\\
\hline
\scriptsize{\textbf{FlowDPS}}

&\scriptsize{24.284}
&\scriptsize{0.657}
&\scriptsize{0.461}
&\scriptsize{0.432}

&\scriptsize{28.507}
&\scriptsize{0.743}
&\scriptsize{0.445}
&\scriptsize{0.293}

\\
\hline
\scriptsize{\textbf{D-Flow}}

&\scriptsize{24.467}
&\scriptsize{0.684}
&\scriptsize{0.488}
&\scriptsize{0.387}

&\scriptsize{27.596}
&\scriptsize{0.709}
&\scriptsize{0.518}
&\scriptsize{0.296}

\\
\hline

\scriptsize{\textbf{FMPlug-W}}

&\scriptsize{\underbar{26.047}}
&\scriptsize{\underbar{0.762}}
&\scriptsize{\underbar{0.369}}
&\scriptsize{0.410}

&\scriptsize{\underbar{30.234}}
&\scriptsize{\underbar{0.813}}
&\scriptsize{\underbar{0.358}}
&\scriptsize{0.367}

\\
\hline
\scriptsize{\textbf{FMPlug-W-R}}

&\scriptsize{\textbf{26.144}}
&\scriptsize{\textbf{0.769}}
&\scriptsize{\textbf{0.355}}
&\scriptsize{\underbar{0.497}}

&\scriptsize{\textbf{30.449}}
&\scriptsize{\textbf{0.821}}
&\scriptsize{\textbf{0.340}}
&\scriptsize{\underbar{0.473}}

\\
\hline

\end{tabular}}
}
\end{center}
\vspace{-1.5em}
\end{table}

\begin{table}[!htpb]
\caption{\textbf{Gaussian Deblur} with additive Gaussian noise ($\sigma = 0.03$). (\textbf{Bold}: best, \underbar{underline}: second best)}
\vspace{-1.5em}
\label{tab:gaussian}
\begin{center}
\setlength{\tabcolsep}{1.0mm}{
\resizebox{\linewidth}{!}{
\begin{tabular}{c c c c c c c c c}
\hline

&\multicolumn{4}{c}{\scriptsize{\textbf{DIV2K~\cite{Agustsson_2017_CVPR_Workshops}}}}
&\multicolumn{4}{c}{\scriptsize{\textbf{AFHQ~\cite{choi2020starganv2}}}}
\\

\cmidrule(lr){2-5}
\cmidrule(lr){6-9}

&\scriptsize{PSNR$\uparrow$}
&\scriptsize{SSIM$\uparrow$}
&\scriptsize{LPIPS$\downarrow$}
&\scriptsize{CLIPIQA$\uparrow$}

&\scriptsize{PSNR$\uparrow$}
&\scriptsize{SSIM$\uparrow$}
&\scriptsize{LPIPS$\downarrow$}
&\scriptsize{CLIPIQA$\uparrow$}

\\
\hline
\scriptsize{\textbf{DIP}} 

&\scriptsize{24.659}
&\scriptsize{0.658}
&\scriptsize{0.469}
&\scriptsize{0.367}

&\scriptsize{28.169}
&\scriptsize{0.680}
&\scriptsize{0.472}
&\scriptsize{0.283}

\\
\hline
\scriptsize{\textbf{FlowChef}} 

&\scriptsize{20.248}
&\scriptsize{0.474}
&\scriptsize{0.629}
&\scriptsize{0.236}

&\scriptsize{24.366}
&\scriptsize{0.661}
&\scriptsize{0.533}
&\scriptsize{\underbar{0.298}}

\\
\hline
\scriptsize{\textbf{FlowDPS}}

&\scriptsize{19.900}
&\scriptsize{0.436}
&\scriptsize{0.615}
&\scriptsize{0.190}

&\scriptsize{24.612}
&\scriptsize{0.626}
&\scriptsize{0.532}
&\scriptsize{0.155}

\\
\hline
\scriptsize{\textbf{D-Flow}}

&\scriptsize{24.798}
&\scriptsize{0.683}
&\scriptsize{0.390}
&\scriptsize{\textbf{0.464}}

&\scriptsize{28.955}
&\scriptsize{0.752}
&\scriptsize{0.458}
&\scriptsize{\textbf{0.316}}

\\
\hline
\scriptsize{\textbf{FMPlug-W}}

&\scriptsize{\underbar{25.995}}
&\scriptsize{\underbar{0.749}}
&\scriptsize{\underbar{0.387}}
&\scriptsize{\underbar{0.413}}

&\scriptsize{\underbar{30.315}}
&\scriptsize{\underbar{0.805}}
&\scriptsize{\underbar{0.378}}
&\scriptsize{0.246}

\\
\hline
\scriptsize{\textbf{FMPlug-W-R}}

&\scriptsize{\textbf{26.125}}
&\scriptsize{\textbf{0.757}}
&\scriptsize{\textbf{0.378}}
&\scriptsize{0.412}

&\scriptsize{\textbf{30.322}}
&\scriptsize{\textbf{0.808}}
&\scriptsize{\textbf{0.372}}
&\scriptsize{0.248}

\\
\hline
\end{tabular}}
}
\end{center}
\vspace{-2em}
\end{table}

We test our FMPlug method against other SOTA methods on two linear IPs: super-resolution and Gaussian debluring. 

\textbf{Datasets, tasks, and evaluation metrics} \quad 
Following \cite{kim2025flowdpsflowdrivenposteriorsampling}, we use two datasets: $25$ DIV2K ~\cite{Agustsson_2017_CVPR_Workshops} validation images and $25$ AFHQ cat ~\cite{choi2020starganv2} images. We set the image resolution to $512 \times 512$ by resizing and cropping the original. We consider two tasks: i) 4$\times$ super-resolution (SR) from $128 \times 128$ to $512 \times 512$; ii) Gaussian deblurring (GD) with a kernel size of $61$ and standard deviation of $3.0$. Both tasks are added with Gaussian noise $\sigma = 0.03$. For metrics, we use PSNR and SSIM for pixel-level difference, and LPIPS for perceptual difference, and CLIPIQA, a no-reference quality metric. 

\textbf{Competing methods} \quad 
We compare our FMPlug (W: warm-up, R: regularized, Number of Function Evaluations (NFE) $=3$) with Deep Image Prior \cite{Ulyanov_2020} (an untrained image prior), D-Flow ($\text{NFE}=3$) \cite{ben2024d} (a SOTA plug-in method), FlowDPS ($\text{NFE}=28$) \cite{kim2025flowdpsflowdrivenposteriorsampling} (a SOTA interleaving method) and FlowChef ($\text{NFE}=28$) \cite{patel2024steering} (another SOTA interleaving method). For a fair comparison, we use the pretrained SD3 model \cite{esser2024scalingrectifiedflowtransformers} as the backbone for all methods that require pretrained priors, i.e., foundation priors. 

\textbf{No text-prompt}
For a fair comparison, we disable text prompt and classifier-free guidance for all methods. It is also not clear how reliably we can generate pertinent text prompts for degraded images. 

\begin{figure*}[!htbp]
\centering
\captionsetup[sub]{font=footnotesize, labelfont=bf, justification=centering}

\begin{subfigure}[b]{0.24\textwidth}
    \includegraphics[width=\linewidth]{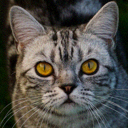}
    \caption{Measurement}
\end{subfigure}%
\hfill 
\begin{subfigure}[b]{0.24\textwidth}
    \includegraphics[width=\linewidth]{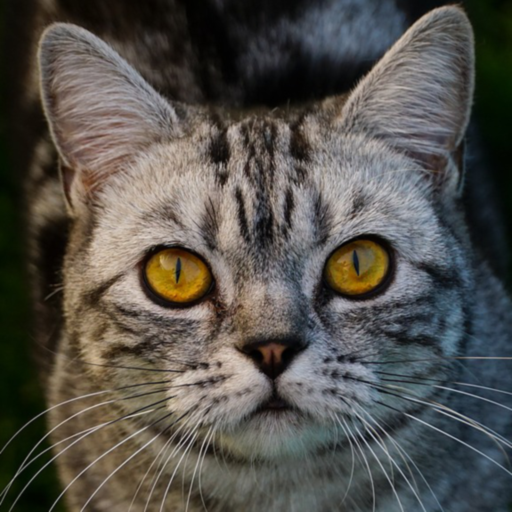}
    \caption{Ground Truth}
\end{subfigure}%
\hfill 
\begin{subfigure}[b]{0.24\textwidth}
    \includegraphics[width=\linewidth]{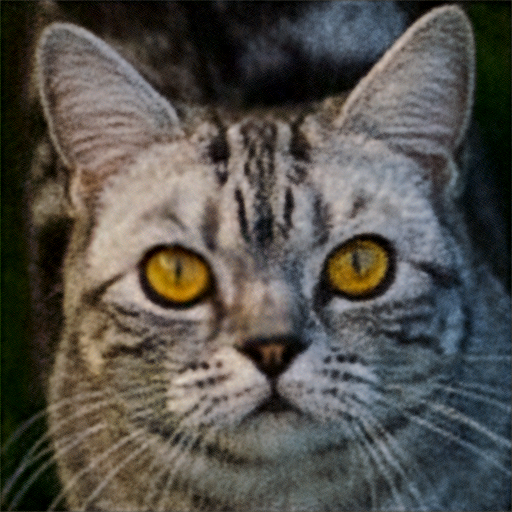}
    \caption{DIP}
\end{subfigure}%
\hfill 
\begin{subfigure}[b]{0.24\textwidth}
    \includegraphics[width=\linewidth]{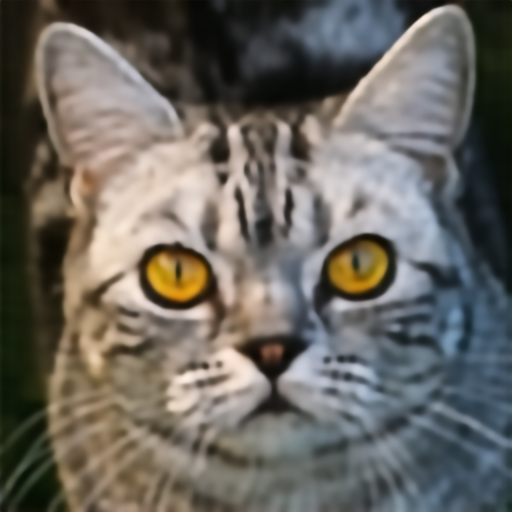}
    \caption{FlowChef}
\end{subfigure}

\vspace{0.5em}

\begin{subfigure}[b]{0.24\textwidth}
    \includegraphics[width=\linewidth]{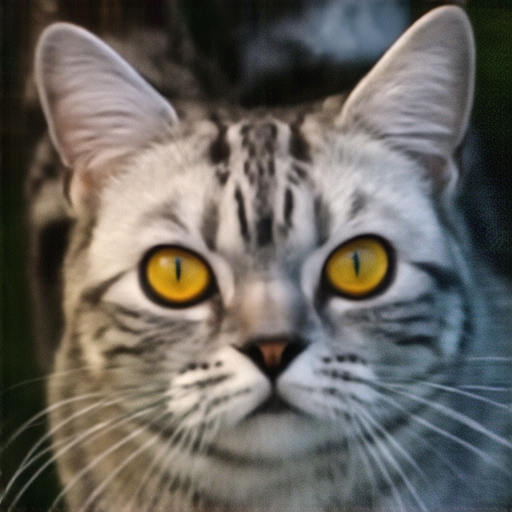}
    \caption{FlowDPS}
\end{subfigure}%
\hfill 
\begin{subfigure}[b]{0.24\textwidth}
    \includegraphics[width=\linewidth]{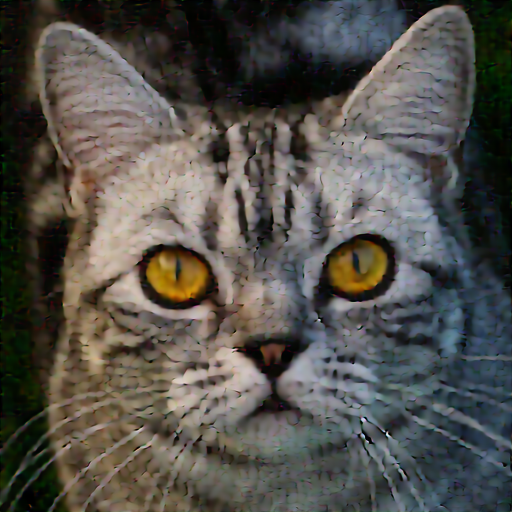}
    \caption{D-Flow}
\end{subfigure}%
\hfill 
\begin{subfigure}[b]{0.24\textwidth}
    \includegraphics[width=\linewidth]{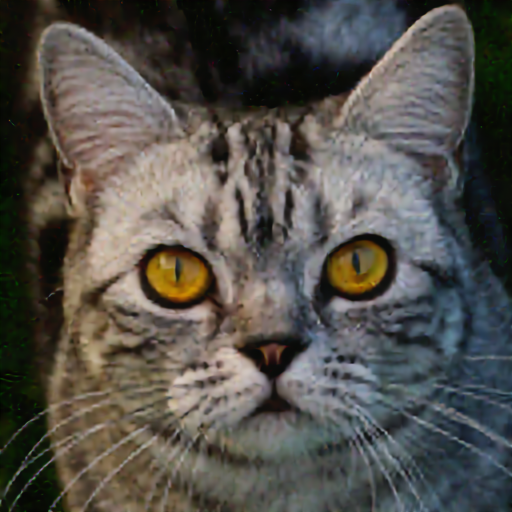}
    \caption{FMPlug-W}
\end{subfigure}%
\hfill 
\begin{subfigure}[b]{0.24\textwidth}
    \includegraphics[width=\linewidth]{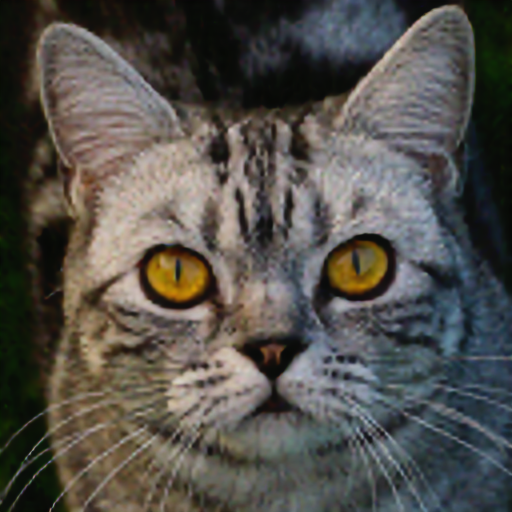}
    \caption{FMPlug-W-R}
\end{subfigure}

\caption{Qualitative results for $4 \times$ super-resolution on the AFHQ dataset.}
\label{fig:visual_superres_afhq}
\end{figure*}

\begin{figure*}[!htbp]
\centering
\captionsetup[sub]{font=footnotesize, labelfont=bf, justification=centering}

\begin{subfigure}[b]{0.24\textwidth}
    \includegraphics[width=\linewidth]{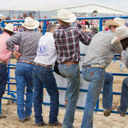}
    \caption{Measurement}
\end{subfigure}%
\hfill 
\begin{subfigure}[b]{0.24\textwidth}
    \includegraphics[width=\linewidth]{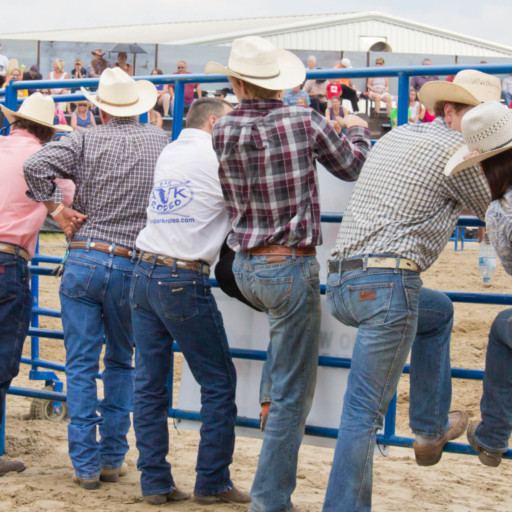}
    \caption{Ground Truth}
\end{subfigure}%
\hfill 
\begin{subfigure}[b]{0.24\textwidth}
    \includegraphics[width=\linewidth]{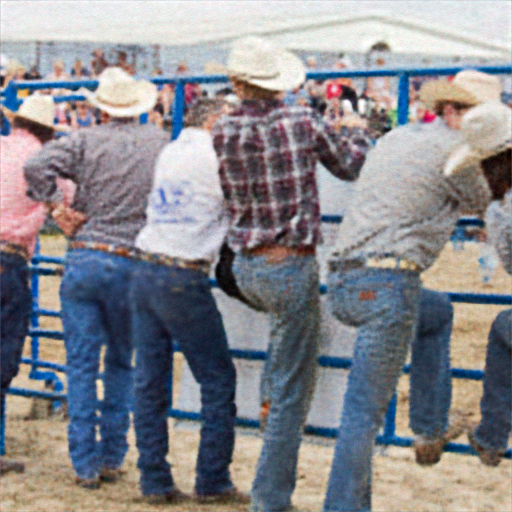}
    \caption{DIP}
\end{subfigure}%
\hfill 
\begin{subfigure}[b]{0.24\textwidth}
    \includegraphics[width=\linewidth]{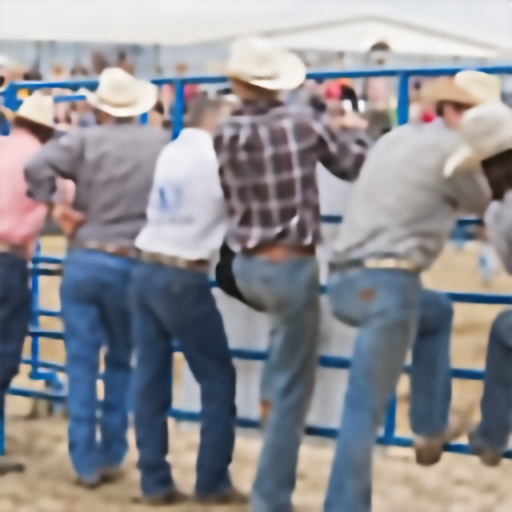}
    \caption{FlowChef}
\end{subfigure}

\vspace{0.5em}

\begin{subfigure}[b]{0.24\textwidth}
    \includegraphics[width=\linewidth]{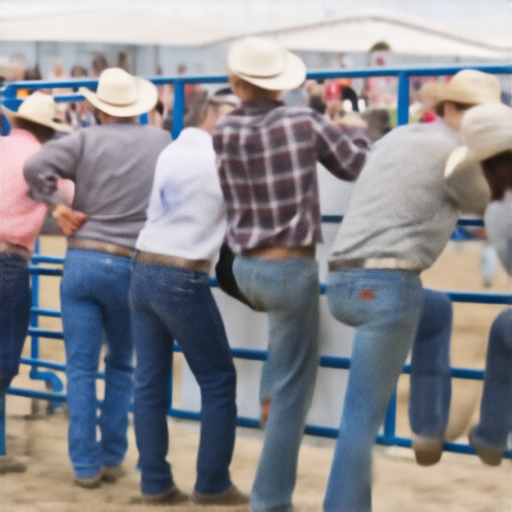}
    \caption{FlowDPS}
\end{subfigure}%
\hfill 
\begin{subfigure}[b]{0.24\textwidth}
    \includegraphics[width=\linewidth]{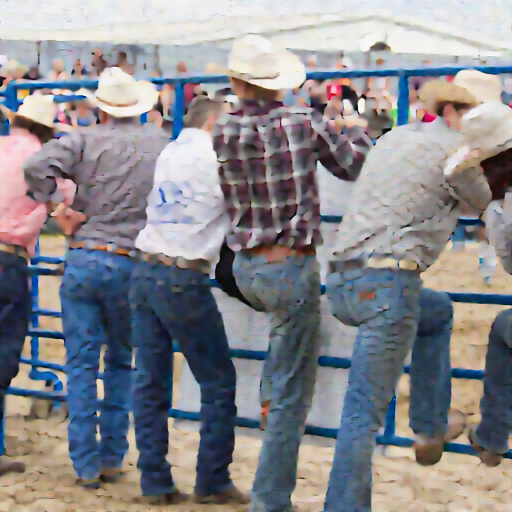}
    \caption{D-Flow}
\end{subfigure}%
\hfill 
\begin{subfigure}[b]{0.24\textwidth}
    \includegraphics[width=\linewidth]{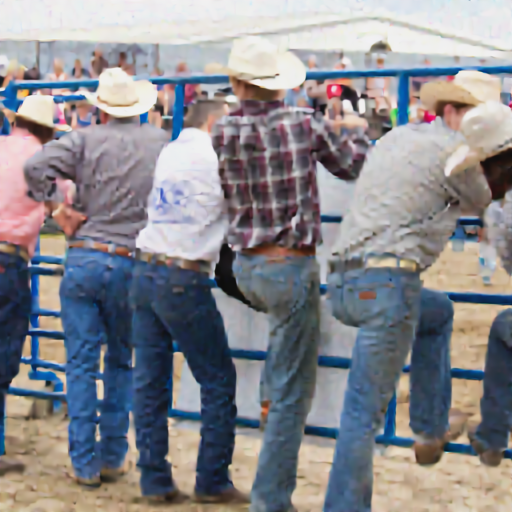}
    \caption{FMPlug-W}
\end{subfigure}%
\hfill 
\begin{subfigure}[b]{0.24\textwidth}
    \includegraphics[width=\linewidth]{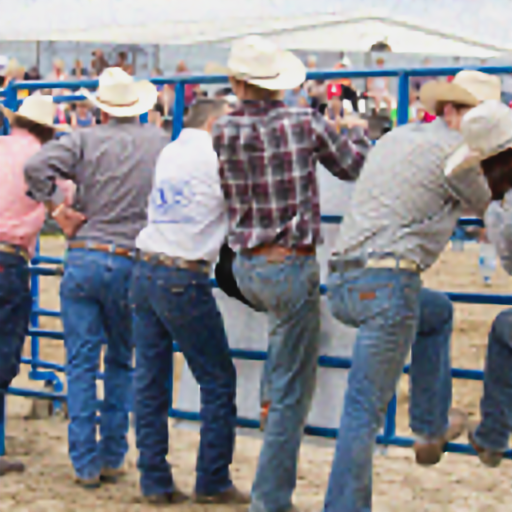}
    \caption{FMPlug-W-R}
\end{subfigure}

\caption{Qualitative results for $4 \times$ super-resolution on the DIV2K dataset.}
\label{fig:visual_superres_div}
\end{figure*}

\begin{figure*}[!htbp]
\centering
\captionsetup[sub]{font=footnotesize, labelfont=bf, justification=centering}

\begin{subfigure}[b]{0.24\textwidth}
    \includegraphics[width=\linewidth]{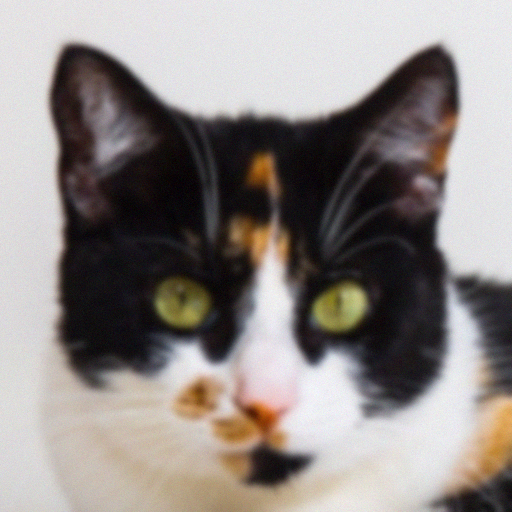}
    \caption{Measurement}
\end{subfigure}%
\hfill 
\begin{subfigure}[b]{0.24\textwidth}
    \includegraphics[width=\linewidth]{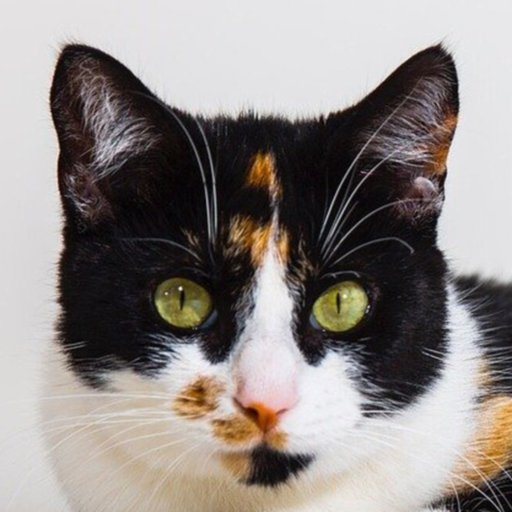}
    \caption{Ground Truth}
\end{subfigure}%
\hfill 
\begin{subfigure}[b]{0.24\textwidth}
    \includegraphics[width=\linewidth]{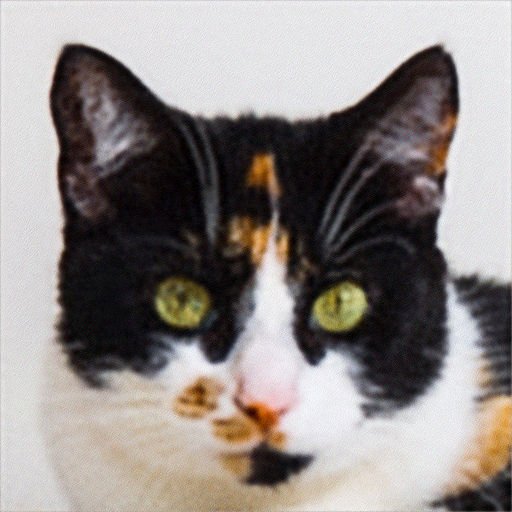}
    \caption{DIP}
\end{subfigure}%
\hfill 
\begin{subfigure}[b]{0.24\textwidth}
    \includegraphics[width=\linewidth]{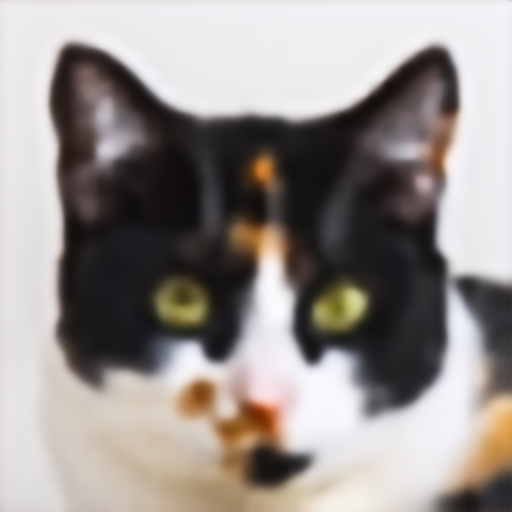}
    \caption{FlowChef}
\end{subfigure}

\vspace{0.5em}

\begin{subfigure}[b]{0.24\textwidth}
    \includegraphics[width=\linewidth]{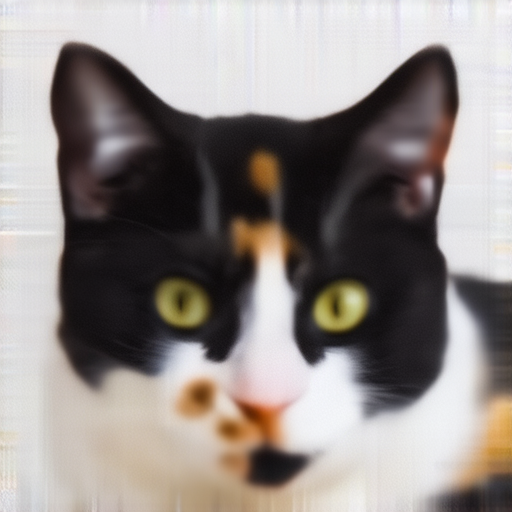}
    \caption{FlowDPS}
\end{subfigure}%
\hfill 
\begin{subfigure}[b]{0.24\textwidth}
    \includegraphics[width=\linewidth]{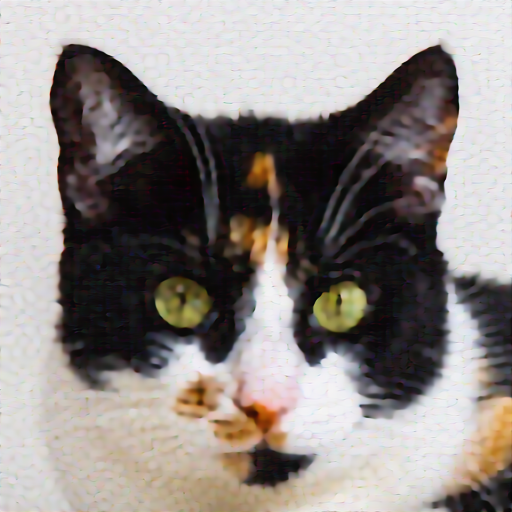}
    \caption{D-Flow}
\end{subfigure}%
\hfill 
\begin{subfigure}[b]{0.24\textwidth}
    \includegraphics[width=\linewidth]{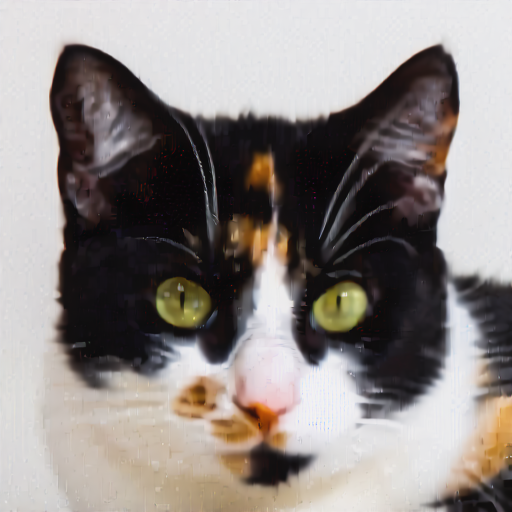}
    \caption{FMPlug-W}
\end{subfigure}%
\hfill 
\begin{subfigure}[b]{0.24\textwidth}
    \includegraphics[width=\linewidth]{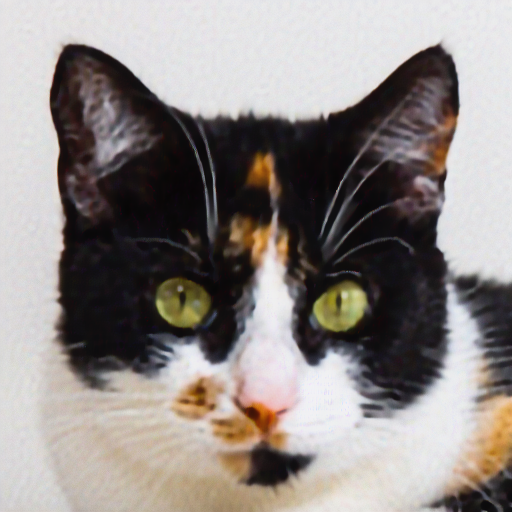}
    \caption{FMPlug-W-R}
\end{subfigure}

\caption{Qualitative results for Gaussian deblurring on the AFHQ dataset.}
\label{fig:visual_gaussian_afhq}
\end{figure*}

\begin{figure*}[!htbp]
\vspace{-1em}
    \centering
    \captionsetup[sub]{font=footnotesize, labelfont=bf, justification=centering}
    \begin{subfigure}[b]{0.24\textwidth}
        \includegraphics[width=\linewidth]{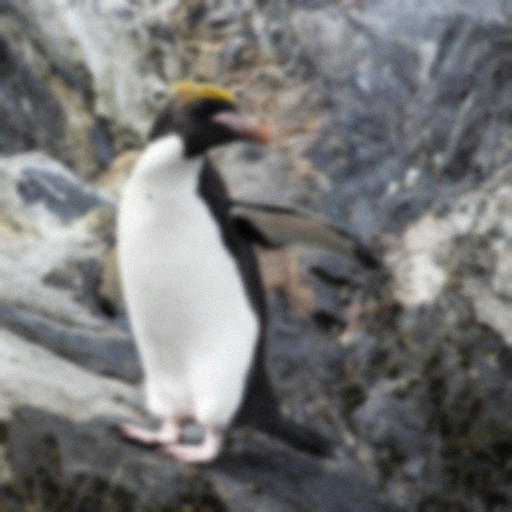}
        \caption{Measurement}
    \end{subfigure}
    \hfill
    \begin{subfigure}[b]{0.24\textwidth}
        \includegraphics[width=\linewidth]{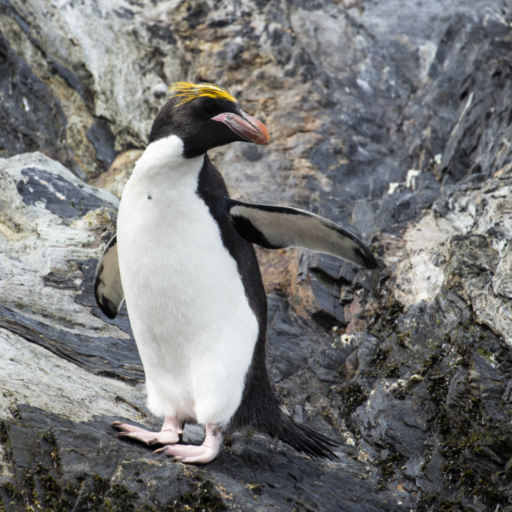}
        \caption{Ground Truth}
    \end{subfigure}
    \hfill
    \begin{subfigure}[b]{0.24\textwidth}
        \includegraphics[width=\linewidth]{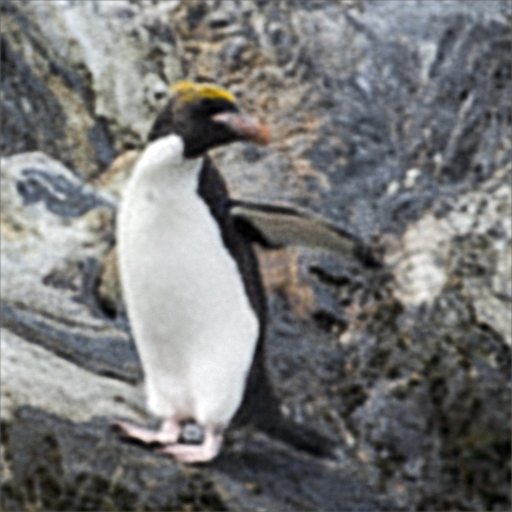}
        \caption{DIP}
    \end{subfigure}
    \hfill
    \begin{subfigure}[b]{0.24\textwidth}
        \includegraphics[width=\linewidth]{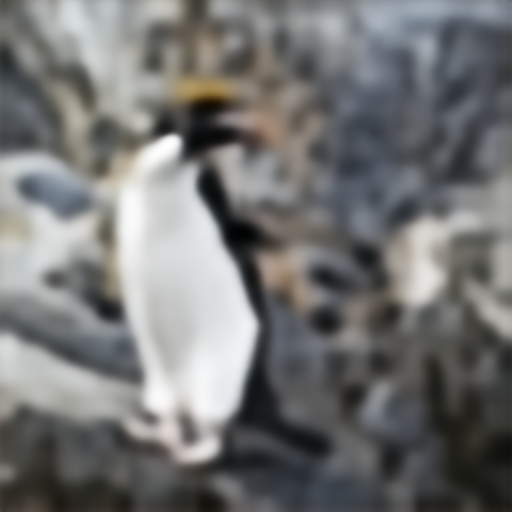}
        \caption{FlowChef}
    \end{subfigure}
    \hfill 
    \\
    \begin{subfigure}[b]{0.24\textwidth}
        \includegraphics[width=\linewidth]{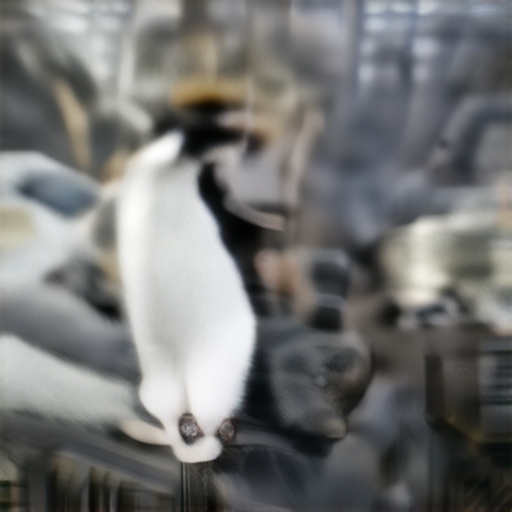}
        \caption{FlowDPS}
    \end{subfigure}
    \hfill
    \begin{subfigure}[b]{0.24\textwidth}
        \includegraphics[width=\linewidth]{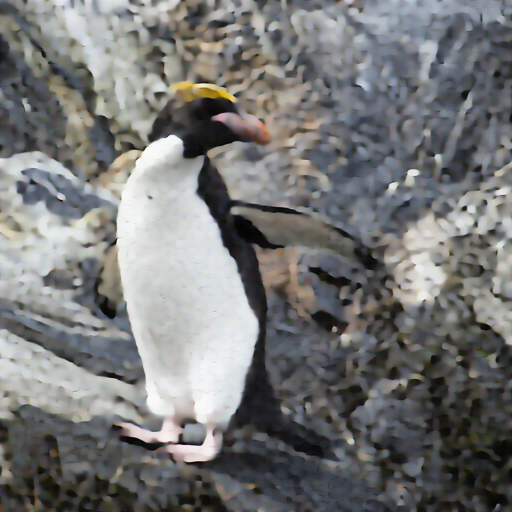}
        \caption{D-Flow}
    \end{subfigure}
    \hfill
    \begin{subfigure}[b]{0.24\textwidth}
        \includegraphics[width=\linewidth]{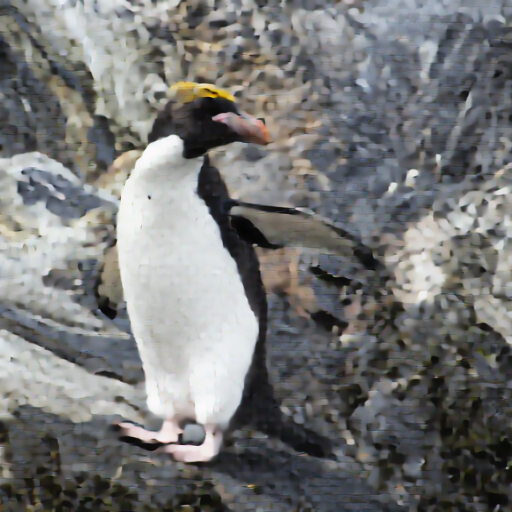}
        \caption{FMPlug-W}
    \end{subfigure}
    \hfill
    \begin{subfigure}[b]{0.24\textwidth}
        \includegraphics[width=\linewidth]{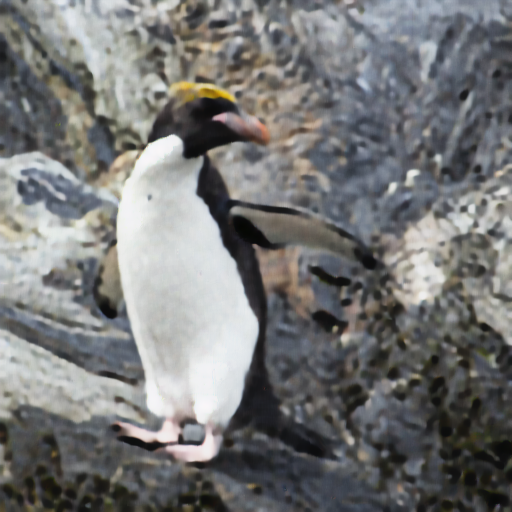}
        \caption{FMPlug-W-R}
    \end{subfigure}
    \caption{Qualitative results for Gaussian deblurring on DIV2K.}
    \label{fig:visual_gaussian_div}
    \vspace{-0.5em}
\end{figure*}

\textbf{Observations} \quad 
\cref{tab:sr4} and \cref{tab:gaussian} summarize the quantitative results. We can observe that: \textbf{(1)} Our FMPlug-W-R is the overall winner by all metrics but CLIPIQA, a no-reference metric. While on the relatively easy SR task the performance gaps between different methods seem minor, on GD the gaps are significant: FlowChef and FlowDPS without prompt guidance lag behind even the untrained DIP by large margins and generate visually blurry and oversmooth images as shown in~\cref{fig:visual_superres_afhq,fig:visual_gaussian_div,fig:visual_gaussian_afhq,fig:visual_superres_div}, suggesting the general struggle of interleaving methods to ensure simultaneous measurement and manifold feasibility; \textbf{(2)} For plug-in methods, our FMPlug-W improves upon D-Flow, by considerable margins based on all metrics but CLIPIQA, showing the solid advantage of our warm-up strategy in \cref{eq:warm} over their initialization strategy in \cref{eq:d-flow-init}; and \textbf{(3)} FMPlug-W-R further improves PSNR and SSIM slightly over FMPlug-W, with the largest improvement seen in CLIPIQA, showing stronger visual quality. This confirms the benefits brought about by the precise Gaussianity regularization in \cref{eq:gauss-proj}.

\bibliographystyle{IEEEtran}
\bibliography{reference}

\end{document}